# General topology of the Universe

Aalok Pandya*

Department of Physics, University of Rajasthan, Jaipur 302004, India.

**Abstract**

General Topology of the universe is described. It is concluded that topology of the present universe is greater or stronger than the topology of the universe in the past and topology of the future universe will be stronger or greater than the present topology of the universe. Consequently, the universe remains unbounded.

___________________________

\* *E- mail address*: belagem@datainfosys.net

A comprehensive discussion on the geometry and topology of the universe, is given by Lachieze-Rey and Luminet [1], and also the global aspects in gravitation and cosmology are described at full length by Pankaj Joshi [3]. Topological aspects of the birth of the universe have been discussed by Miao Li [4]. And topological aspects of the large-scale structures in the universe, have been discussed by Hartle and Hawking [5], Ellis and Hawking [6] and others [7-8]. But, these attempts were by and large focused on the discussion of geometry of the manifold of the universe. Author feels that apart from the manifold description, other general topological properties and the topological changes following them are equally important. The general topology of the universe can be uniquely described if its physical picture and features are specified. A topological space is an abstract conception and thus it may not be always possible to attribute physical properties to it. But, the other way round is certainly possible i.e. topology of a physical object with evident metricity i.e. distance features inbuilt, can certainly be described. We have several models of the universe at the threshold of scientific development, but we consider only one scenario wherein the universe is the single largest entity, which encloses every physical object that exists in nature including space and time. The discussion in this letter is developed on two logical grounds such that firstly it justifies the observed facts and secondly it satisfies the mathematical conditions of the whole space topology. The metric of the universe can be considered as the super metric or the universal metric defined on the whole space $U$. Then such a universal topology is the set of all possible sets and subsets including the null set and excluding nothing. With the consideration that everything that exists in nature is contained in the universe, the complement of the universe is obviously nonexistent.

By definition universal topology being collection of all possible sets and subsets, is always discrete in nature. Thus the topology of the universe is a discrete topology. Also, the whole space $U$ is always open as well as closed. We explain this geometrical duality in this context using both geometrical as well as physical explanation. Since $U \cap \{\phi\} = \{\phi\}$; $U$ is closed if its complement $\sim U$ is open and $U$ is open if its complement $\sim U$ is closed, as the null space $\phi$ is both open as well as closed.

Mathematically, if the closure of the space $U$ is given by $\bar{U}$, the condition of the whole space implies that: $U = \bar{U}$.

Hence, if the closure of the universe is equal to the universe itself then the topology of the universe is said to be closed. Now, we look for the frontier points and the limit points of the universe. If the universe is not static and is continuously expanding, its farthest possible extents can be easily approximated. The radius of the expanding universe can be given by the well-known relation $R = \frac{c}{H}$, or alternatively $R \approx c\tau$; where $H^{-1}$ is Hubble age, $c$ is the velocity of light, and $\tau$ is characteristic time scale or the age of the universe. Alternatively, for the sake of generality, this distance can also be given by $L_{Horizon}$ [1] with a good precision as the (comoving) radius of the observable universe (with or without inflation): $L_{Horizon} \approx \frac{u}{(1-\gamma)} L_{Hubble}$ ;

or $L = c \int_{t=t_r}^{t=t_o} \frac{dt}{R(t)}$.

This follows from the logic that the size of the universe can be determined by means of the distance traversed by the fastest moving particle ever since the genesis of the universe. Also, the universe is considered to be homogeneous and isotropic. Obviously, this radius is only instantaneous radius and is increasing continuously. Any point on the farthest end is frontier point of the universal metric. Since this frontier point is moving away or receding with the speed of light, no other object or point can ever overtake it, and this frontier point remains the frontier point. For the points at the frontier of the universe moving with the velocity $c$, no limit or stop over exists, and therefore the limit points of the universe lie within the universe. And the metric of the universe having its limit points within itself implies a closed topology. Moreover, the frontier points will always be at frontier and interior point will always be interior. This is how we justify by both physical and geometrical arguments that the universe is topologically closed. However, the metric of the universe is closed but it cannot be bounded. With the expansion of the universe, the continuous inclusion of more number of space (or space-time) points restricts the universe from being bounded and compact.

If the universe goes on expanding then the topology of the universe will tend to be greater or stronger topology. The topology of the universe at any point of time in future will be stronger or greater than the present topology, and it will further tend to be greater and stronger. This is due to the fact that it will have more and more number of open sets, following this enhancement. If we consider the topology of the universe defined on the whole space $(U, \Im)$, at some point of time in the past then we can construct the topologies of the universe at present and in the future, relative to this.

The topology of the present universe can be given as:

$\Im_{present} = \{U, \phi, \cup \{x_\mu\}\}$; where $\cup \{x_\mu\}$ is the collection of all space-time points accumulated in $U$, after having defined the topology $\Im_{past}$.

Similarly, we can define the topology of the universe of future as:

$\Im_{future} = \{U, \phi, \cup \{x_\nu\}\}$, where $\cup \{x_\mu\} \subset \cup \{x_\nu\}$.

Hence $\Im_{past} \subset \Im_{present} \subset \Im_{future}$.

In other words $U(past)$ is smaller or weaker than the $U(present)$, and $U(future)$ will be greater or stronger than the $U(present)$. This further guarantees that the universe will remain unbounded. This is in agreement with the statement that universe has no boundary, because a boundary represents in a sense the 'edge' of the universe not detected by any astronomical observations [3].

An important mathematical corollary [9] asserts that boundary of a space can be given by: $\text{Bd}\, U = \overline{U} - U$.

But, since $\overline{U} = U$, the boundary of the space vanishes as $\text{Bd}\, U = \phi$.

And $\text{Bd}\, U = \phi \Leftrightarrow U$ is both open and closed. This agrees with the conclusion, which we have already drawn but by another reasoning. Furthermore, if $d$ is a metric defined on space $U$, a subset $A$ of $U$ is said to be bounded if there is some number $N$ such that

$d(a_1, a_2) \leq N$; for every pair of points $a_1, a_2$ of $A$.

If $A$ is bounded, the diameter of $A$ is defined to be the number

$diam\, A = $ least upper bound $\{d(a_1, a_2) : a_1, a_2 \in A\}$.

But, we know that for $U$ itself, which is continuously expanding, this condition cannot be fulfilled as for every pair $a_1, a_2$; the distance $d(a_1, a_2)$ cannot remain less than or equal to a fixed number $N$ forever.

Also, we can alternatively explain the topological transition of the universe from a weaker or smaller to a stronger or greater topology in terms of matter-energy distribution. Considering the total energy of the universe to be conserved, the continuous expansion of the universe will lead to increase in the intergalactic distances and decline in the matter density over the space. This implies that the topology of the universe in past $U(past)$ was coarser than the $U(present)$, and $U(future)$ will be finer than the $U(present)$.

The quantum state of the universe was described by Hartle and Hawking [5] by the wave function of the universe, $\Psi(h_{ij}, \varphi) = \int d[g_{\mu\nu}] d[\varphi] \exp(-I)$, with the amplitude of the metric $h_{ij}$ and matter fields $\varphi$. As rightly pointed out by Miao LI [4], there exists a problem of how to define the measure of the integral. By its characteristics, the wave function is continuous everywhere and vanishes at the infinity. No box or sphere, exist with finite size for the bound states of this wave function and therefore it implies only unbounded universe. It is argued [4] that if the laws of Physics are parity broken, as found in weak interaction, all four manifolds must be orientable, supporting this argument by a mathematics theorem which asserts that the boundary of an orientable manifold is orientable. Ironically, parity violation is not a universal phenomenon. Also, it is clear in the present discussion by all possible explanations that universe is unbounded. Thus by no means it is possible to prove that the manifold of the universe is orientable.

One can define a manifold over a topology if that topological space is locally homeomorphic to $\mathbf{R}^n$, which means it is Hausdorff and possesses metric properties, and connectedness is well described over that. Moreover, a smooth manifold can be described only if continuity and differentiability is well defined on it. The manifold of the universe can only be described, if it satisfies these requisite conditions, and we know that it does satisfy.

However, the explanation of connectedness and continuity in the context of matter-energy in the universe is a complicated task. But, if the origin of the universe is considered to have followed a single event say Big-Bang, then all the physical objects and events in the universe are causally connected. Thus, one can say that the metric of the space-time is connected too. Physically interesting known space-time examples such as the Schwarzschild geometry and Robertson-Walker space-times are topologically Hausdorff. Also, For the sake of the causal regularity of space-time, the continuum of the space-time structure is necessary which would ensure a globally well-behaved space-time [3]. In nut-shell, at least for the Minkowski space-time, the manifold is globally Euclidean with the topology of $\mathbf{R}^4$ over the Lorentzian metric. In case of general theory of relativity, the principles of GTR effectively imply that it is the space-time metric, and the quantities derived from it, that must appear in the equations for physical quantities and that these equations must reduce to the flat space-time case when the metric is Minkowskian. This is the basic content of the GTR where the space-time manifold is allowed to have topologies other than $\mathbf{R}^4$ too. However, globally, there could be important differences in the causal structure due to a different space-time topology, strong gravitational fields and so on [3].

Even though the manifold of the universe can have different possible topologies depending upon the physical conditions considered in a specific scenario, the universe being the whole space as described in this letter, the global aspects of the general topology of the universe remain unchanged. Thus, this suggestion is resurfaced that the mathematical conditionality of the whole space topology can be used as a constraint to describe the topology of the universe.

## Acknowledgement

Author wishes to thank Dr. T. R. Sheshadri, Prof. Pankaj Joshi, and Prof. N. K. Sharma for their suggestions and encouragement.

# References

[1] M. Lachieze-Rey and J. P. Luminet, Phys. Rep. **254** (1995) 135-214; M. Lachieze-Rey and and J. P. Luminet, gr-qc/9605010.

[2]. Neil J. Cornish, David N. Spergel, and Glenn D. Starkman, astro-ph/9708083.

[3] Pankaj S. Joshi, Global Aspects in Gravitation and Cosmology, Oxford Science Publications (1993).

[4] Miao Li, Phys. Lett. **B** 173 (1986) 229.

[5] J. B. Hartle and S. W. Hawking, Phys. Rev. **D** 28 (1983) 2960; S. W. Hawking, Nucl. Phys. **B** 39 (1984) 257.

[6] S. W. Hawking and G. F. R. Ellis, The Large Scale Structure of Space-Time, Cambridge University Press, Cambridge (1973).

[7] Adrian L. Melott, Phys. Rep. 193 (1990).

[8] M. J. Duncan and Lars Gerhard Jensen, Is the Universe Euclidean? CERN-TH.5311/89.

[9] J. R. Munkres, Topology: A First Course, Prentice-Hall Inc. (1975); D. Bushaw, Elements of General Topology, John Wiley & Sons Inc. (1963).